\documentstyle[12pt,sw20lart]{article}

\input tcilatex
\QQQ{Language}{
American English
}

\begin{document}

\author{Nikolai Laskin\thanks{%
E-mail: nlaskin@rocketmail.com}}
\title{{\bf Stochastic Theory of \ Foreign Exchange Market Dynamics }}
\date{Isotrace Laboratory, University of Toronto\\
60 St. George Street, Toronto, ON M5S 1A7\\
Canada }
\maketitle

\begin{abstract}
A new stochastic theory of a foreign exchange markets dynamics is developed.
As a result we have the new probability distribution which well describes
statistical and scaling dependencies ''experimentally'' observed in foreign
exchange markets in recent years. The developed dynamic theory is compared
with well-known phenomenological Levy distribution approach which is widely
applied to this problem. It is shown that the developed stochastic dynamics
and phenomenological approach based on the Levy distribution give the same
statistical and scaling dependencies.

{\it PACS }number(s): 02.50 Ey, 05.40. Fb
\end{abstract}

\section{Introduction}

Statistical behavior of the foreign exchange (FX) markets and price
fluctuations in currency have been the subject of studies in recent years 
\cite{Mantega}, \cite{Dodge}. The high-frequency data for financial markets
has made it possible to investigate market dynamics on timescales as short
as 1 min, a value close to the minimum time needed to perform transaction in
the market. It was observed that the short-term price fluctuations in FX
market, for example, between US dollar and German mark, has the same
statistical behavior as the velocity differences in hydrodynamic turbulence 
\cite{Dodge}.

Probability distributions in turbulence well fit the experimental data by
superposition of the Gaussians with log-normal distributions of its
variances \cite{Dodge},\cite{Castaing}. The convergence of the velocity
differences distributions toward a Gaussian shape corresponds to a decrease
of the log-normal variance for increasing of spatial distances $\Delta r$.
In the FX fluctuation dynamics the statistical distributions of the price
difference separated by time $\Delta t$ was elaborated by the theoretical
model of a Levy walk or Levy flight \cite{Mantega},\cite{Levy}.

I have developed a new stochastic dynamic theory for the FX market. The
Langevin type stochastic differential equation is proposed. I introduce the
random force with some general characteristics in order to fit the
theoretical predictions of my model with observed statistical dependencies.

It is shown that the developed model describes exactly the ''experimental''
data of the dynamics of a price index of the New York Stock Exchange. The
probability distributions of the Standard \& Poor's 500 index differences $%
\Delta x$ are reproduced exactly by the $P_{Laskin}(\Delta x,\Delta t)$. The
new model describes well the scaling behavior of the ''probability of
return'' $P_{\Delta t}(0)$ as a function of $\Delta t$ (for definitions, see 
\cite{Mantega}).

\section{Dynamic model of price changes}

We will describe the dynamics of the FX price $x(t)$ by the stochastic
differential equation

\begin{equation}
\stackrel{\cdot }{x}(t)=F(t),  \label{eq1}
\end{equation}

where $F(t)$ is the random force. The quantity which were interested about
is the price differences separated by the time scale $\Delta t$,

\begin{equation}
\Delta x=x(t+\Delta t)-x(t),  \label{eq2}
\end{equation}

The distribution function of stochastic process $\Delta x$ is defined as
follow

\begin{equation}
P(\Delta x,\Delta t)=<\delta (\Delta x-\int\limits_t^{t+\Delta t}d\tau
F(\tau ))>,  \label{eq3}
\end{equation}

where \TEXTsymbol{<}...\TEXTsymbol{>} means the averaging over the all
possible realizations of the random force $F(t)$. Let us construct the
general stochastic force $F(t)$ by the following way

\begin{equation}
F(t)=\sum\limits_{k=1}^na_k\varphi (t-t_k),  \label{eq4}
\end{equation}

here $a_k$ are the random amplitudes, $\varphi (t)$ is the response
function, $t_k$ are the homogeneously distributed (on time interval $[0,T]$)
moments of time, the number $n$ of which obeys the Poisson law.

Thus, the averaging includes three statistically independent averaging
procedures:

1. Averaging over random amplitudes $a_k$, $<...>_{a_k}$,

\begin{equation}
<...>_{a_k}=\int da_1...da_nP(a_1,...,a_n)...,  \label{eq5}
\end{equation}

where $P(a_1,...,a_n)$ is the probability distribution of amplitudes $a_k$.

2. Averaging over $t_k$ on time interval $T$,

\begin{equation}
<...>_T=\frac 1T\int\limits_0^Tdt_1...\frac 1T\int\limits_0^Tdt_n....
\label{eq6}
\end{equation}

3. Averaging over random numbers $n$ of time moments $t_k$,

\begin{equation}
<...>_n=\sum\limits_{n=0}^n\frac{\overline{n}^n}{n!}e^{-\overline{n}}...,
\label{eq7}
\end{equation}

where $\overline{n}=\nu T$ and $\nu $ is the density of points $t_k$ on time
interval $T$.

Taking into account the definition Eq.(\ref{eq3}) and performing the
averaging in accordance with Eqs.(\ref{eq5})-(\ref{eq7}) we will have

\begin{equation}
P_{Laskin}(\Delta x,\Delta t)=\frac 1\pi \int\limits_0^\infty d\xi \cos (\xi
\cdot \Delta x)e^{-L(\xi ,\Delta t)},  \label{eq8}
\end{equation}

where we put notation

\begin{equation}
L(\xi ,\Delta t)=\nu \delta \int\limits_0^{\frac{\sigma \delta }{\sqrt{2}}%
(1-e^{-\Delta t/\delta })}\frac{du}u(1-e^{-u^2}).  \label{eq9}
\end{equation}

For simplicity the Eq.(\ref{eq8}) is considered for the market currency
situation when $P(a_1,...,a_n)$ is factorized as follow

\begin{equation}
P(a_1,...,a_n)=\prod\limits_{k=1}^nP_1(a_k),  \label{eq10}
\end{equation}

where the Gaussian distribution $P_1(a)$ is given by

\begin{equation}
P_1(a)=\frac 1{\sigma \sqrt{2\pi }}\exp (-\frac{a^2}{2\sigma ^2}).
\label{eq11}
\end{equation}

We also choose the response function in the form

\begin{equation}
\varphi (t)=e^{-|t|/\delta }.  \label{eq12}
\end{equation}

The response function describes the influence of a piece of information
which has become available at the delay time $t$ on the decision of a trader
to propose or accept a price change. The parameter $\delta $ is the
character scale of the time delay. We also have considered the response
function in the form

\begin{equation}
\varphi (t)=\frac 1{1+(|t|/\delta )^\beta },  \label{eq13}
\end{equation}
where $\beta $ is a new parameter.

For example, for $\beta =1$ we have

\begin{equation}
P_{Laskin}(\Delta x,\Delta t;\beta =1)=\frac 1\pi \int\limits_0^\infty d\xi
\cos (\xi \cdot \Delta x)e^{-L_\beta (\xi ,\Delta t;\beta =1)},  \label{eq14}
\end{equation}

where

\begin{equation}
L_\beta (\xi ,\Delta t;\beta =1)=\nu \Delta t\int\limits_0^{\ln \frac{\Delta
t+\delta }\delta }\frac{dze^z}{(e^z-1)^2}(1-e^{-\frac{\sigma ^2\delta ^2\xi
^2}2z^2}).  \label{eq15}
\end{equation}

Thus, we derive the new $P_{Laskin}(\Delta x,\Delta t)$ distribution (see
Eqs. (\ref{eq8}), (\ref{eq9})) starting from the differential stochastic
equation Eq.(\ref{eq1}).

As it was mentioned the Levy stable distribution is widely applied to this
problem.

The comparison between $P_{Laskin}(\Delta x,\Delta t)$ transformed to the
form

\begin{equation}
P_{Laskin}(\Delta x,\Delta t)=\frac 1\pi \int\limits_0^\infty dy\cos (y\cdot
\Delta x)\times ,  \label{eq16}
\end{equation}

\[
\times \exp \left\{ -\frac 12D\int\limits_0^{\tau ^2(\Delta t)y^2}\frac{dz}%
z(1-e^{-z})\right\} , 
\]

the Levy stable distribution \cite{Mantega}, \cite{Levy}, \cite{Feller}

\begin{equation}
P_{Levy}(\Delta x,\Delta t)=\frac 1\pi \int\limits_0^\infty dy\cos (y\cdot
\Delta x)e^{-\gamma \Delta ty^\alpha },  \label{eq17}
\end{equation}

and the Gauss distribution

\begin{equation}
P_{Gauss}(\Delta x)=\frac 1{\sqrt{2\pi }\sigma }\exp \{-x^2/2\sigma ^2\}.
\label{eq18}
\end{equation}

is shown in Fig.1.

Figure 2 is a comparison of scaling properties of $P_{Laskin}(0,\Delta t)$,
and $P_{Levy}(0,\Delta t).$

It is interesting that the {\it Laskin} stochastic dynamics and
phenomenological approach based on the {\it Levy} distribution give the same
scaling for the probability of return $P_{\Delta t}(0)$ of the S\&P 500
index variations as a function of the $\Delta t$ \cite{Mantega}.

\section{General FX rate dynamic model}

The main goal of the developed theory is to predict the behavior of currency
exchange market in the future and based on these prediction to develop the
efficient currency market strategy.

As we will see further it is useful to generalize the Eqs.(\ref{eq1}), (\ref
{eq4}) in following way

\begin{equation}
\stackrel{\cdot }{x}(t)=-\lambda x+\sum\limits_{k=1}^na_k\varphi (t-t_k)
\label{eq19}
\end{equation}

where $\lambda $ is a ''price dissipative coefficient'' the financial mean
of which will be discussed further.

Let us define the new PDF $P_{Laskin}(x,t)$ as follow

\begin{equation}
P_{Laskin}(x,t)=<\delta (x-x(t,x_0))>_{a,T,n}  \label{eq20}
\end{equation}

where $x(t,x_0)$ is the formal solution of the Eq.(\ref{eq19}). and

\[
<...>_{a,T,n}=\int da_1...da_nP(a_1,...,a_n)\sum\limits_{n=0}^n\frac{%
\overline{n}^n}{n!}e^{-\overline{n}}\frac 1T\int\limits_0^Tdt_1...\frac
1T\int\limits_0^Tdt_n... 
\]

Using the Eqs.(\ref{eq19}), (\ref{eq20}) it is easy to obtain the evolution
equation for PDF $P_{Laskin}(x,t)$

\begin{equation}
\frac{\partial P_{Laskin}(x,t)}{\partial t}=\lambda \frac \partial {\partial
x}xP_{Laskin}(x,t)+  \label{eq21}
\end{equation}

\[
+\nu \int daP_1(a)\int\limits_{t_0}^tdt^{\prime }\varphi (t-t^{\prime
})\left\{ e^{-a\int\limits_{t_0}^td\tau e^{\lambda (t-\tau )}\varphi (\tau
-t^{\prime })\frac \partial {\partial x}}-1\right\} P_{Laskin}(x,t) 
\]

The initial condition for this equation has the form

\begin{equation}
P_{Laskin}(x,t)=<\delta (x-x(t,x_0))>_{a,T,n}  \label{eq22}
\end{equation}

The Eq.(\ref{eq21}), (\ref{eq22})will serve as the main equations to predict
the behavior of FX market in the future.

It is easy to see that the solution of the problem Eqs.(\ref{eq21}) and (\ref
{eq22}) can be written as

\begin{equation}
P_{Laskin}(x,t)=\frac 1{2\pi }\int\limits_{-\infty }^\infty d\xi e^{i\xi
(x-e^{\lambda (t-t_0)}x_0)}\times  \label{eq23}
\end{equation}

\[
\times \exp \left\{ -\nu \int\limits_{t_0}^tdt^{\prime }\left( 1-W(\xi
\int\limits_{t_0}^td\tau e^{\lambda (t-\tau )}\varphi (\tau -t^{\prime
}))\right) \right\} 
\]

where $W(\xi )$ is characteristic function of the random amplitude $a$. It
is well-known \cite{Feller} that the PDF $P_1(a)$ and characteristic
function $W(\xi )$ are connected each other

\[
P_1(a)=\frac 1{2\pi }\int\limits_{-\infty }^\infty d\xi e^{i\xi a}W(\xi ) 
\]

and

\[
W(\xi )=\int\limits_{-\infty }^\infty dae^{-i\xi a}P_1(a) 
\]

The Eqs.(\ref{eq21}) and (\ref{eq22}) allow one to study the evolution
problems in FX currency market. Some analytically solvable evolution
problems will be demonstrated in the next publication.

\section{Conclusions}

The new dynamic stochastic theory of FX currency market dynamics is
developed. We have established the new {\it Laskin} distribution which well
describes the observed statistical dependencies of the price difference
separated by time scale $\Delta t$ and the {\it Laskin} PDF which allows
elaborate evolution of the currency market and predict behavior of market
dynamics.

The theoretical predictions based on the {\it Laskin} distribution are
compared with phenomenological the {\it Levy} distribution based approach
and observed statistical dependencies. It is interesting that the {\it Laskin%
} stochastic dynamics and phenomenological approach based on the {\it Levy}
distribution give the same statistical and scaling dependencies.

\section{Figure captions}

Fig.1. Comparison between $P_{Laskin}(\Delta x,\Delta t)$, $P_{Levy}(\Delta
x,\Delta t)$ and $P_{Gauss}(\Delta x)$ (see the definitions Eqs.(\ref{eq16}%
), (\ref{eq17}), (\ref{eq18})). For $P_{Laskin}(\Delta x,\Delta t)$ the
parameters are $\Delta t=0.5$, $D=1.4$, $S=0.16$ and $\tau (\Delta t)=S\cdot
(1-e^{-\Delta t})$. For $P_{Levy}(\Delta x,\Delta t)$ the parameters are $%
\alpha =0.00375$ and $\gamma =1.4$. For $P_{Gauss}(\Delta x)$, $\sigma
=0.0508$ (see, \cite{Mantega}).

Fig.2 Scaling dependencies of $P_{Laskin}(0,\Delta t)$ and $%
P_{Levy}(0,\Delta t)$ ''probabilities of return''. All parameters are the
same as for Fig.1.

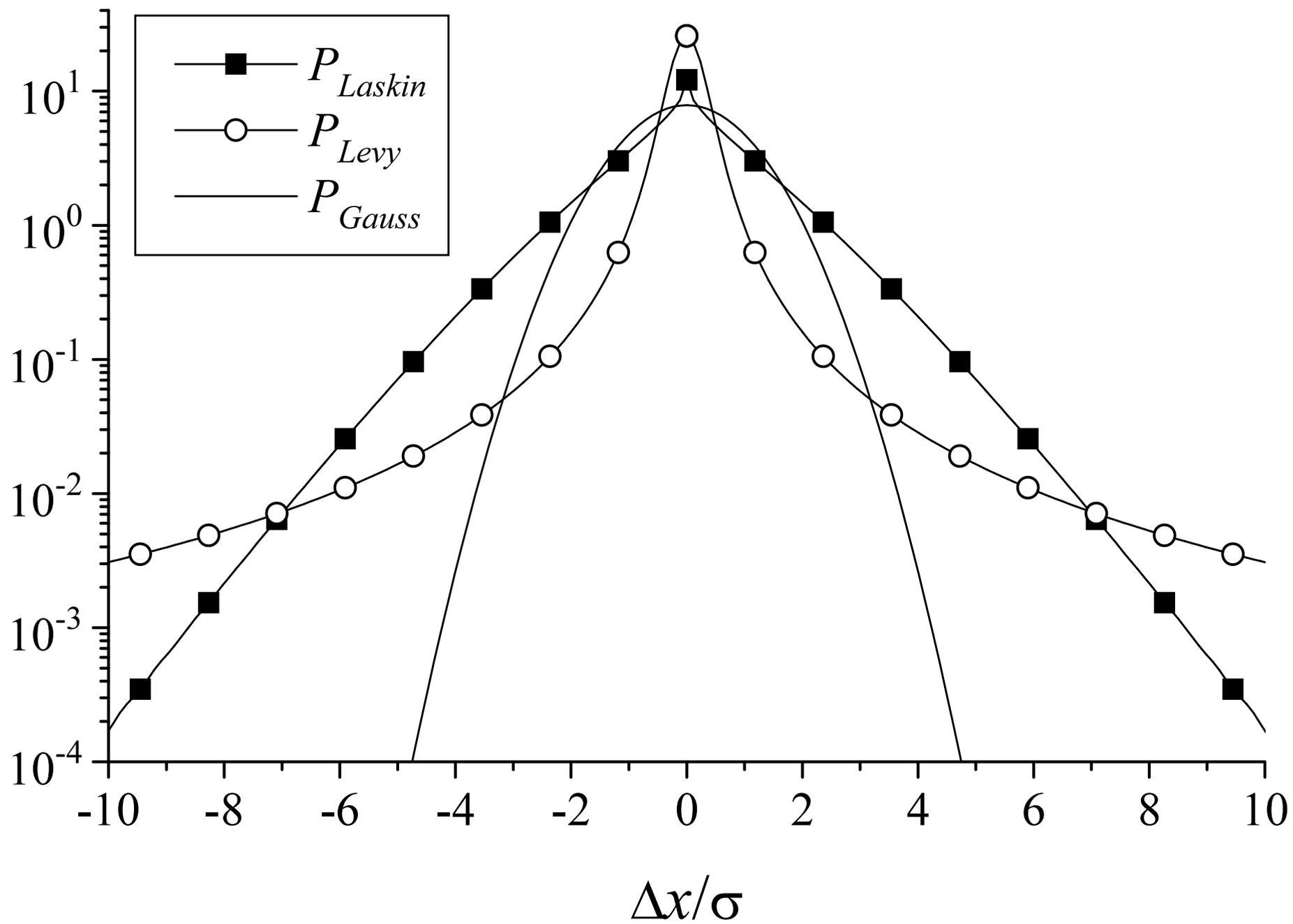
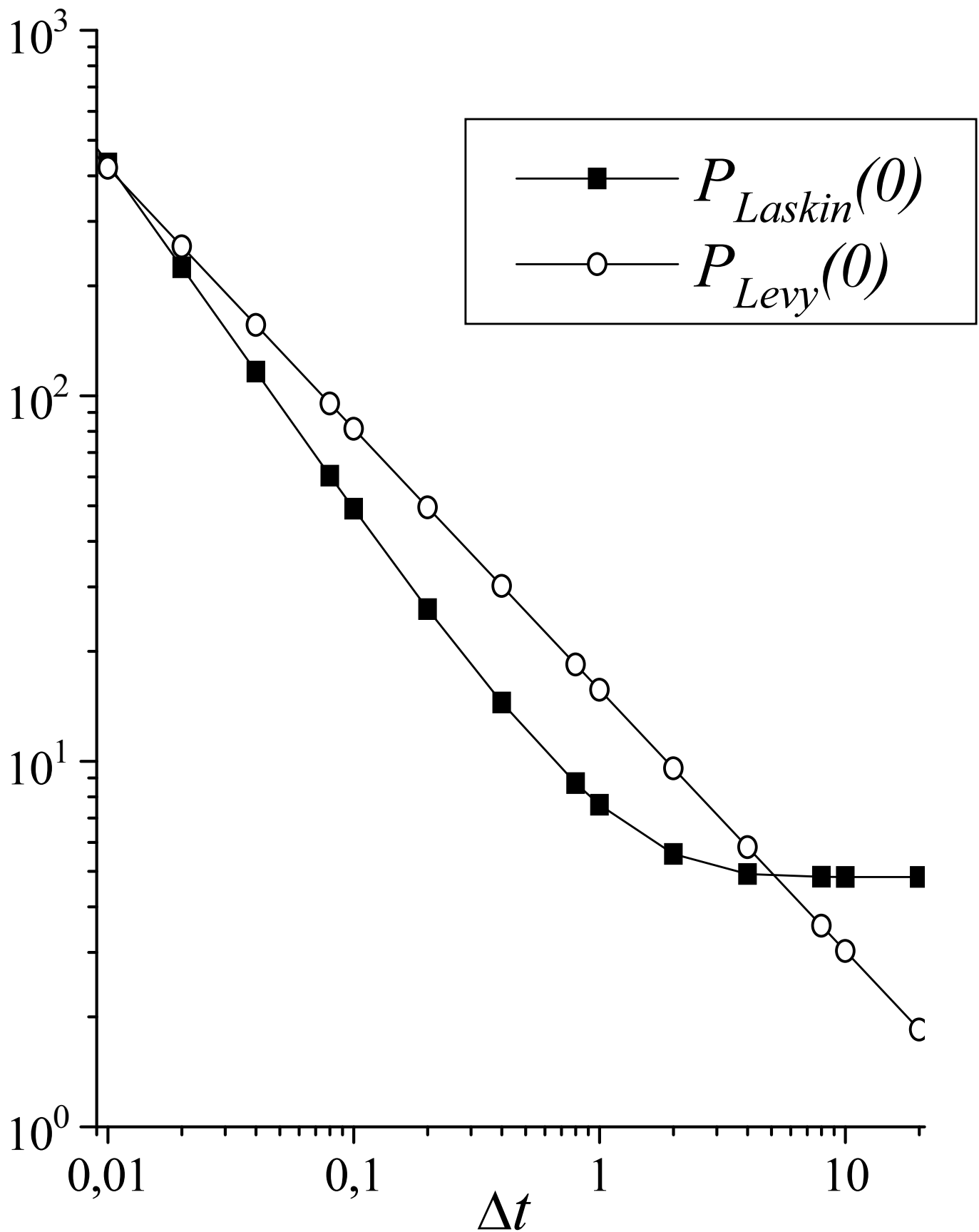
\end{document}